Scan-less full-field fluorescence-lifetime dual-comb microscopy using two-dimensional spectral mapping and frequency multiplexing of dual-optical-comb beats

Takahiko Mizuno,[1,2] Eiji Hase,[1,2] Takeo Minamikawa,[1-3] Yu Tokizane,[1] Ryo Oe,[4] Hidenori Koresawa,[4] Hirotsugu Yamamoto,[2,3,5] and Takeshi Yasui[1-3,*]

[1]Institute of Post-LED Photonics (pLED), Tokushima University, 2-1 Minami-Josanjima, Tokushima, Tokushima 770-8506, Japan

[2]JST, ERATO, MINOSHIMA Intelligent Optical Synthesizer Project, 2-1 Minami-Josanjima, Tokushima, Tokushima 770-8506, Japan

[3]Graduate School of Technology, Industrial and Social Sciences, Tokushima University, 2-1 Minami-Josanjima, Tokushima, Tokushima 770-8506, Japan

[4]Graduate School of Advanced Technology and Science, Tokushima University, 2-1 Minami-Josanjima, Tokushima, Tokushima 770-8506, Japan

[5]Center for Optical Research and Education, Utsunomiya University, 7-1-2, Yoto, Utsunomiya, Tochigi 321-8585, Japan

Correspondence and requests for materials should be addressed to

yasui.takeshi@tokushima-u.ac.jp.

# Abstract

Fluorescence lifetime imaging microscopy (FLIM) is a powerful tool for quantitative fluorescence imaging because fluorescence lifetime is independent of concentration of fluorescent molecules or excitation/detection efficiency and is robust to photobleaching. However, since FLIM is based on point-to-point measurements, mechanical scanning of a focal spot is needed for forming an image, which hampers rapid imaging. In this article, we demonstrate scan-less full-field FLIM based on a one-to-one correspondence between two-dimensional (2D) image pixels and frequency-multiplexed RF signals. A vast number of dual-optical-comb beats between dual optical frequency combs is effectively adopted for 2D spectral mapping and high-density frequency multiplexing in radio-frequency region. Bimodal images of fluorescence amplitude and lifetime are obtained with high quantitativeness from amplitude and phase spectra of fluorescence RF comb modes without the need for mechanical scanning. The proposed method will be useful for rapid quantitative fluorescence imaging in life science.

Fluorescence microscopy[1] is a powerful tool for observing the localization and migration of specific molecules and proteins in cells and tissue by molecular labelling with fluorescent probes or proteins. Although fluorescence intensity is a representative imaging modality in fluorescence microscopy, it is often influenced by not only the concentration of fluorescent molecules but also photobleaching and excitation/detection efficiency. Therefore, the absolute fluorescence intensity often lacks quantitativeness. Although the fluorescence intensity ratio between two different fluorescence or excitation wavelengths[2,3] is an alternative method that can compensate for the drawbacks associated with absolute fluorescence intensity, it is influenced by differences in focal position between the two wavelengths.

An interesting imaging modality for quantitative fluorescence microscopy is fluorescence lifetime, which is used in fluorescence lifetime imaging microscopy (FLIM)[4]. The fluorescence lifetime is determined by the types of fluorescent molecules and the environment surrounding the molecules, and shows little dependence on the concentration of fluorescent molecules, photobleaching, and excitation/detection efficiency. Therefore, the fluorescence lifetime is more quantitative than the fluorescence intensity. It is also sensitive to changes in the micro-environment around fluorescent molecules, such as temperature[5] or pH[6]. Furthermore, even though the signals from two fluorescent molecules with similar fluorescence wavelengths are spectrally overlapped, molecular selectivity can be achieved based on different fluorescence lifetimes[7].

Fluorescence lifetime can be measured by the time-correlated single-photon-counting method (TC-SPC) with pulsed excitation[4,8,9] or the phase measurement (PM) method with sinusoidal excitation[4,10,11]. Both methods are based on point measurement, and they have advantages and disadvantages, including data acquisition time. Therefore, mechanical scanning of the focal point is necessary for obtaining an image using fluorescence lifetime. Such mechanical scanning often limits the frame rate of FLIM. While use of a rotating polygonal scanning mirror[12] or a combined microlens array and pinhole array[13] can boosts the frame rate in fluorescence-intensity microscopy, such fast mechanical scanning is not compatible with the TC-SPC or PM method. FLIM without the need for mechanical scanning, that is to say, scan-less full-field FLIM, is promising for rapid fluorescence lifetime imaging and will expand the range of applications of FLIM in life science. Since the PM method can achieve a higher data acquisition rate than the TC-SPC method, a promising approach is to parallelize the PM method without the need for mechanical scanning. Scan-less wide-field FLIM was achieved by combining the PM method with a camera[14,15]; however, the low detection efficiency of the camera hinders rapid image acquisition. Another scan-less wide-field FLIM was demonstrated by using a combination of the PM method and single-pixel imaging[16]. However, the need for consecutive excitation of different spatially-coded light patterns in this method hampers rapid image acquisition.

One interesting approach expandable to scan-less full-field FLIM is fluorescence imaging by radiofrequency-tagged emission (FIRE)[17-19], which is based on a combination of frequency multiplexing and spectral mapping in the radio-frequency (RF) region. In this method, a 1D array of focal spots with different RF modulation frequencies is generated by digitally synthesized optical beating with an acousto-optic deflector, and is then radiated onto a sample. After passing through a confocal slit, the radiated fluorescence is detected by a point photodetector. Since the detected fluorescence signal has a one-to-one correspondence between RF beat frequencies and line pixels via frequency multiplexing and spectral mapping, a line fluorescence intensity image is reconstructed from a group of frequency multiplexed

RF beat signals without the need for mechanical scanning of the focal point. However, the limited number of frequency-multiplexing RF beats (typically, a few hundred) makes it difficult to expand this approach to scan-less 2D images. Therefore, 1D mechanical scanning of the focal line is still required to obtain a 2D image. If one can increase the number of frequency-multiplexed RF beats up to its square, scan-less full-field fluorescence imaging, namely, 2D-FIRE, could be achieved by using a combination of ultra-dense frequency multiplexing and 2D spectral mapping.

One potential method for ultra-dense frequency-multiplexed RF signals is to use an optical frequency comb (OFC)[20-23]. An OFC is composed of a large number of optical frequency spikes (typically, tens to hundreds of thousands) with a constant frequency spacing (= $f_{rep}$), and functions as an optical carrier with a vast number of discrete channels. Furthermore, multi-frequency heterodyning interferometry between dual OFCs (frequency spacing = $f_{rep1}$, $f_{rep2}$) with slightly different frequency spacing (= $\Delta f_{rep} = f_{rep2} - f_{rep1}$), namely, dual-optical-comb beating, generates a secondary optical beat comb in the RF region, namely, an RF comb. Since an RF comb is a replica of an OFC whose frequency scale is reduced by $f_{rep1}/\Delta f_{rep}$, it has been used for broadband high-precision spectroscopy, for example, in dual-comb spectroscopy (DCS)[24-26]. On the other hand, because the number of RF comb modes is equal to that of OFC modes, it is sufficient for frequency multiplexing RF signals in 2D-FIRE. 2D spectral mapping of OFC modes can be achieved by a combination of a virtually imaged phased array (VIPA)[27,28] and a diffraction grating. Such 2D spectral mapping has been effectively used for not only optical wavelength division demultiplexing[29] and broadband spectroscopy[30] but also for 2D imaging[31,32]. Recently, 2D spectral mapping has been combined with DCS for scan-less full-field confocal microscopy based on a one-to-one correspondence between OFC modes and 2D image pixels[33]. However, the application of this dual-comb microscopy (DCM) is limited to coherent light, such as reflected or transmitted light, due to the use of multi-frequency-heterodyning interferometry.

In this article, we expand DCM into scan-less full-field FLIM or 2D-FIRE, that is, so-called f-DCM, by combining ultra-dense frequency-multiplexed RF signals of dual-optical-comb beats with 2D spectral mapping using VIPA and a diffraction grating. Based on the one-to-one correspondence between dual-optical-comb RF beats and 2D image pixels, a 2D image of fluorescence intensity is obtained from the amplitude spectrum of ultra-dense frequency-multiplexed fluorescence RF signals without the need for mechanical scanning of the focal spot. Furthermore, 2D imaging of fluorescence lifetime is demonstrated based on parallel phase measurement of fluorescence RF signals, which corresponds to parallelization of the PM method.

## Results

Figures 1(a) and 1(b) respectively illustrate an experimental setup and a sample of f-DCM to implement the 2D spectral mapping (Fig. 2) and frequency multiplexing (Fig. 3) of dual-optical-comb beats, which is described in the Methods section together with details on the experimental and analytical methodology employed for the following measurements.

**Scan-less imaging of fluorescence intensity**

We first demonstrated scan-less imaging of fluorescence intensity. A Rhodamine 6G aqueous solution (molar concentration = 300 µM, excitation wavelength = 461–533 nm, fluorescence wavelength = 543–617 nm, fluorescence lifetime = 4.08 ns)[34] was filled into the chamber to serve as the fluorescent sample. We measured the region around the Group6/Element1 pattern (line spacing = 39 µm/lp) of the target. Figure 4(a) shows a temporal waveform of fluorescence RF comb modes from the sample (sampling interval = 97 fs, number of accumulated signals = 100,000), in which the effective time and lab time are given as the lower (window size = 9.96 ns) and upper scales (window size = 1.02 ms). The effective time and lab time are related by a factor of $f_{rep1}/\Delta f_{rep}$ (= 100,385,982/978 = 102,644). The temporal behaviour of the fluorescence RF comb modes was observed as a centre burst. The Fourier transform of the temporal waveform gives the amplitude and phase spectra of the fluorescence RF comb modes. Figure 4(b) shows the amplitude spectrum of the fluorescence RF comb modes. As many as 44,400 high-density distributed RF modes appeared, as shown by the filled region inside the spectrum. The spectral envelope reflects the 2D distribution of fluorescence intensity in the sample. Then, we reconstructed the fluorescence intensity image from the amplitude spectrum based on the one-to-one correspondence between RF comb modes and 2D image pixels. Figure 4(c) shows the reconstructed fluorescence amplitude image (image size = 234 μm by 79 μm, pixel size = 300 pixels by 148 pixels). The fluorescence amplitude image of the target pattern was confirmed to have high image contrast.

We next investigated the dependence of the signal-to-noise ratio (SNR) on the number of accumulated signals in the fluorescence intensity imaging. Figure 5 shows fluorescence intensity images when the number of accumulated signals was set to (a) 100, (b) 1,000, (c) 10,000, and (d) 100,000, respectively. The image contrast changed depending on the number of accumulated signals. For a more quantitative analysis, we defined image SNR as the ratio of the mean to the standard deviation of the fluorescence intensity for a region of interest (ROI) in the image. Figure 5(e) shows the relation between the number of accumulated signals and image SNR. The image SNR linearly increased depending on the square root of the number of accumulated signals, indicating that shot-noise-limited measurement was achieved.

**Scan-less imaging of fluorescence lifetime**

We performed scan-less bimodal imaging of fluorescence lifetime and fluorescence intensity for a different region of the same sample as in the experiment above. Figure 6(a) shows a temporal waveform of fluorescence RF comb modes from the sample (window size of effective time = 9.96 ns, sampling interval = 97 fs, number of accumulated signals = 100,000). By performing Fourier transformation of the temporal waveform, we obtained the amplitude and phase spectra of fluorescence RF comb modes as shown in Figs. 6(b) and 6(c). Then, we reconstructed two kinds of image, namely, in fluorescence intensity and fluorescence phase delay, based on the one-to-one correspondence between RF comb modes and 2D image pixels, as shown in Figs. 6(d) and 6(e) (image size = 234 μm by 79 μm, pixel size = 300 pixels by 148 pixels). Finally, we obtained the fluorescence lifetime image of the sample by substituting the phase-delay spectrum of fluorescence RF comb modes for Eq. (1), as shown in Fig. 6(f). The test pattern was clearly confirmed in both images. However, it is important to note that the mechanism of the image contrast is different between them. The mean and standard deviation of fluorescence lifetime in the image region were 4.1 ns and 0.7 ns, respectively, which agree well with the lifetime value reported in the literature for this fluorescence solution (= 4.08 ns)[34].

To evaluate the quantitativeness of fluorescence lifetime imaging, we acquired fluorescence lifetime images of samples with different fluorescence lifetimes. The corresponding fluorescence lifetime images are shown in Fig. 7: (a) Rhodamine 6G aqueous solution, (b) Rhodamine B aqueous solution, (c) Rhodamine B methanol solution, and (d) Rhodamine B ethanol solution. While all images showed similar patterns of the test chart, the lifetime values were significantly different among them: (a) 4.1±0.7 ns, (b) 1.8±0.4 ns, (c) 2.4±0.6 ns, and (d) 3.0±0.5 ns. Figure 7(e) shows a correlation between the literature values[34,35] and the measured values of fluorescence lifetime. A high correlation value (= 0.96) was obtained between them. When the measurement accuracy is defined as a root mean square error between measured values and literature values, the measurement accuracy of 0.07 ns was achieved in this experiment, indicating the high quantitativeness of fluorescence lifetime imaging using f-DCM.

## Discussion

We first discuss the possibility of further increasing the image acquisition rate with the proposed f-DCM method. In principle, the maximum rate of image acquisition is equal to $\Delta f_{rep}$ because the temporal waveform of image-encoded fluorescence RF comb modes is acquired at a rate of $\Delta f_{rep}$. In the present setup, $\Delta f_{rep}$ was set to 997 Hz; in other words, single-shot imaging can be achieved at 997 Hz. However, the scan-less full-field fluorescence imaging demonstrated above needs a large number of accumulated images to obtain a high-SNR image (see Fig. 5). The reason why image accumulation is needed is the considerably weak excitation light, which is limited by the efficiency of nonlinear wavelength conversion in the PPLN crystal and signal loss in the 2D spectral disperser. Considering the total power of excitation light (= 242 μW) at the focus and the number of 2D image pixels (= 149 × 200 = 29,800 pixels), the excitation power per pixel was only 8 nW, which is much lower than that in the conventional PM method (typically, µW order)[10,11] or FIRE[17-19]. Since the image SNR in the proposed system was shot-noise-limited, as shown in Fig. 5(e), a higher total power of excitation light will efficiently contribute to improvement of the image SNR or reduction of the image acquisition time.

We next discuss the possibility of achieving confocality in the f-DCM. Confocal fluorescence microscopy[36] is a powerful tool for low-invasive two-dimensional (2D) optical sectioning or three-dimensional (3D) imaging in life science because it provides confocality, which enables depth selectivity on micrometre order, together with stray light elimination, in thick biological specimens. In usual confocal fluorescence microscopy, confocality is achieved by placing a detector pinhole at a conjugate position with respect to a focused excitation spot in a specimen. In the case of f-DCM, since a 2D array of focal spots is formed in the sample, one has to place a 2D array of pinholes at a conjugate position with respect to the 2D array of focal spots. One possible approach for implementing a 2D pinhole array with flexibility is to use a spatial light modulator (SLM). Work is in progress to install an SLM as a flexible 2D pinhole array into f-DCM.

In summary, we demonstrated f-DCM for scan-less full-field fluorescence lifetime imaging and fluorescence intensity imaging. Use of dual-optical-comb beats between dual OFCs enables us to expand the DCM[33] to fluorescence microscopy. The scan-less full-field imaging capability of f-DCM was achieved by 2D spectral mapping and frequency multiplexing of dual-optical-comb beats. The maximum acquisition rate of fluorescence lifetime imaging was 997 Hz, although image accumulation was required to obtain a high SNR. The measurement accuracy of 0.07 ns in fluorescence lifetime measurement was ensured in the scan-less imaging by employing a parallelized PM method with high-density multiplexed RF comb modes. We discussed the possibility of further reducing the image acquisition time of f-DCM and the compatibility with confocal fluorescence microscopy. f-DCM will be a powerful tool for rapid quantitative fluorescence microscopy.

## Methods

**2D spectral mapping of dual-optical-comb beats**

When an OFC (OFC1, frequency spacing = $f_{rep1}$, optical frequency = $\nu_i = \nu_1$, $\nu_2$, $\nu_3$, ⋯, $\nu_n$) beam is spatially overlapped with another OFC beam with a slightly different frequency spacing (OFC2, frequency spacing = $f_{rep2} = f_{rep1} + \Delta f_{rep}$), a group of optical beats is generated as dual-optical-comb beats (beat frequency = $b_i = i\Delta f_{rep}$ = $\Delta f_{rep}$, $2\Delta f_{rep}$, $3\Delta f_{rep}$, ⋯, $n\Delta f_{rep}$) via multi-frequency-heterodyning interference between adjacent modes of OFC1 and OFC2, as shown in the upper part of Fig. 1(a). In the RF region, such dual-optical-comb beats show a large number of frequency spikes with different frequencies $b_i$ (= $\Delta f_{rep}$, $2\Delta f_{rep}$, $3\Delta f_{rep}$, ⋯, $n\Delta f_{rep}$), namely, an RF comb, as shown in the lower part of Fig. 1(a). Since the RF comb is regarded as a replica of OFC1 in which the frequency spacing is downscaled from $f_{rep1}$ to $\Delta f_{rep}$, a one-to-one correspondence is established between OFC modes $\nu_i$ and RF comb modes $b_i$. In other words, such dual-optical-comb beating is equivalent to tagging each OFC mode $\nu_i$ with a unique RF beat frequency $b_i$.

2D spectral mapping of OFC modes is achieved by a 2D spectral disperser, composed of a VIPA and a diffraction grating[33]. The 2D spectral disperser enables us to spatially develop OFC modes as a 2D spectrograph of zigzag lines depending on the optical frequency or wavelength, as shown in Fig. 1(b), enabling another one-to-one correspondence between 2D image pixels pi and OFC modes $v_i$. When beams of both OFC1 and OFC2 are simultaneously fed into the 2D spectral disperser, each adjacent mode pair of dual-optical-comb beats is spatially overlapped and forms a common focal point at different positions depending on their optical frequency.

**Scan-less 2D imaging with radio-frequency-multiplexed fluorescence signals**

2D spectral mapping of dual-optical-comb beats enables us to combine (i) the one-to-one correspondence between OFC modes $v_i$ and RF comb modes $b_i$ with (ii) the one-to-one correspondence between 2D image pixels $p_i$ and OFC modes $v_i$. This results in a one-to-one correspondence between 2D image pixels $p_i$ and RF comb modes $b_i$ at the fluorescence excitation side, as shown in Figs. 2(a) and 2(b). The resulting dual-optical-comb beats with high-density frequency multiplexing are suitable for a frequency-multiplexed excitation source in 2D-FIRE. When 2D focal points of dual-optical-comb beats with different RF modulation frequencies excite a fluorescent sample, the fluorescence appears from the corresponding 2D spots with equivalent modulation frequencies as a fluorescence RF comb. At the fluorescence detection side, 2D image pixels pi have a one-to-one correspondence with fluorescence RF comb modes $b_i$, as shown in Figs. 2(c) and 2(d). Therefore, by detecting the frequency-multiplexed fluorescence signal with a single photodetector and reconstructing a 2D image based on the one-to-one correspondence between fluorescence 2D image pixels $p_i$ and fluorescence RF comb modes $b_i$, a full-field 2D fluorescence image is acquired without the need for mechanical scanning of the focal spot.

**Bimodal imaging of fluorescence intensity and lifetime**

Let us consider the case where excitation light with modulation frequency $b_i$ is incident on the sample, as shown in Fig. 8(a). Although the resulting fluorescence is also modulated synchronously with the excitation light modulated at $b_i$, there is a difference of amplitude and phase between them, as shown in Fig. 8(b). The fluorescence intensity image can be obtained by parallel measurement of amplitude values of the modulated fluorescence for all modulation frequencies corresponding to 2D image pixels, that is, measurement of the amplitude spectrum of the fluorescence RF comb modes [see Fig. 2(d)]. On the other hand, the phase delay $\phi_i$ of the modulated fluorescence from the modulated excitation light is related to the fluorescence lifetime $\tau_i$ of the sample by

$$\tau_i = \frac{\tan\phi_i}{2\pi b_i} \tag{1}$$

Therefore, measurement of the phase delay spectrum between the excitation RF comb and the fluorescence RF comb enables scan-less full-field fluorescence lifetime imaging.

**Experimental setup**

Figure 1(a) shows a schematic diagram of the experimental setup. We used a pair of custom-built femtosecond Er-fiber OFCs (OFC1, centre wavelength = 1560 nm, spectral bandwidth = 15 nm, mean output power = 167 mW, $f_{ceo1}$ = 21.4 MHz, $f_{rep1}$ = 100,385,982 Hz; OFC2, centre wavelength = 1560 nm, spectral bandwidth = 15 nm, mean output power = 121 mW, $f_{ceo2}$ = 21.4 MHz, $f_{rep2}$ = $f_{rep1}$ +$\Delta f_{rep}$ = 100,386,960 Hz, $\Delta f_{rep}$ = 978 Hz). Near-infrared light from OFC1 and OFC2 was converted into green light via a combined wavelength-conversion process of second-harmonic-generation (SHG) and sum-frequency-generation (SFG) in chirped periodically-poled lithium-niobate (PPLN) crystals (HC Photonics Corp., Hsinchu, Taiwan, TY-0299, operating wavelength range = 1540–1570 nm). We obtained dual-optical-comb beats around 520 nm by spatially overlapping the two green light beams with powers of 11.1 mW and 9.7 mW from OFC1 and OFC2, as shown in the inset in Fig. 1(a). The green dual-optical-comb beats were fed into a 2D spectral disperser, composed of a VIPA (Light Machinery, Inc., Nepean, Ontario, Canada, OP-6721-6743-2, free spectral range = 14.8 GHz, finesse = 65) and a diffraction grating (Thorlabs Inc., Newton, NJ, USA, GR25-1205, groove density = 1200 grooves/mm, blaze wavelength = 500 nm). A 2D spectrograph of the green dual-optical-comb beats was formed at the optical Fourier plane of the grating and was then relayed onto a sample as a 2D array of focal points of excitation light by a combination of lenses (L1 and L2) and a dry-type objective lens (OL, Nikon Corp., Tokyo, Japan, CFI Plan Fluor 10×, numerical aperture = 0.30, working distance = 16 mm). Fluorescence from the sample was reflected by a dichroic mirror (DM, Semrock Inc., Lake Forest, IL, USA, FF562-Di03-25x36, transmission wavelength = 569.5–950 nm, reflection wavelength = 350–554.5 nm), was filtered by an optical bandpass filter (BPF, Semrock Inc., Lake Forest, IL, USA, FF01-593/40-25, passband = 573–613 nm), and was then detected with a photomultiplier (PMT, Hamamatsu Photonics K.K., Hamamatsu, Japan, H7844, sensitive wavelength range = 185–900 nm, quantum sensitive efficiency at 590 nm = 10 %). The temporal waveform of the detected electrical signal was acquired by a digitizer (National Instruments Corp., Austin, Texas, USA, NI PCI-5122, sampling rate = $f_{rep2}$ =

100,386,960 samples/s, number of sampling points = 102,644, resolution = 14 bit). The Fourier transform of the acquired temporal waveform gives the amplitude spectrum and phase spectrum of the fluorescence RF comb. Finally, images of fluorescence amplitude and lifetime were obtained based on the one-to-one correspondence between fluorescence RF comb modes and fluorescence image pixels.

**Sample**

As a fluorescent sample, we combined a 1951 USAF high-resolution target (Edmund Optics Inc., Barrington, NJ, USA, #55-622, negative type, size = 2 inch by 2 inch, minimum line spacing = 1.55 μm/lp) with a fluorescent solution. Figure 1(b) shows a schematic drawing of the sample. We used the target as a spatial mask and put a chamber containing a fluorescent solution behind the target. The 2D array of focal spots of dual-optical-comb beats comes from the target side, passes through the transparent region of the target, and excites fluorescence in the chamber. The fluorescence radiates in all directions from the fluorescent solution, and its back-propagating component passes through the target and thus contains information about the target in the form of a target image. We prepared four kinds of fluorescent solutions with different lifetimes, as shown in Table 1.

**Data availability**

The data that support the findings of this study are available from the corresponding author upon reasonable request.

**Funding**

Exploratory Research for Advanced Technology (ERATO), Japan Science and Technology Agency (MINOSHIMA Intelligent Optical Synthesizer Project, JPMJER1304); Japan Society for the Promotion of Science (18H01901, 18K13768, 19H00871); Cabinet Office, Government of Japan (Subsidy for Reg. Univ. and Reg. Ind. Creation); Nakatani Foundation for Advancement of Measuring Technologies in Biomedical Engineering.

**Author contributions**

T.Y. and H.Y. conceived the project. Tak.Miz. and E.H. performed the experiments and/or analysed the data. Tak.Miz. and T.Y. wrote the manuscript. Tak.Min., Y.T, R.O., and H.K. discussed the results and commented on the manuscript.

**Competing financial interests statement**

The authors declare no competing financial interests.

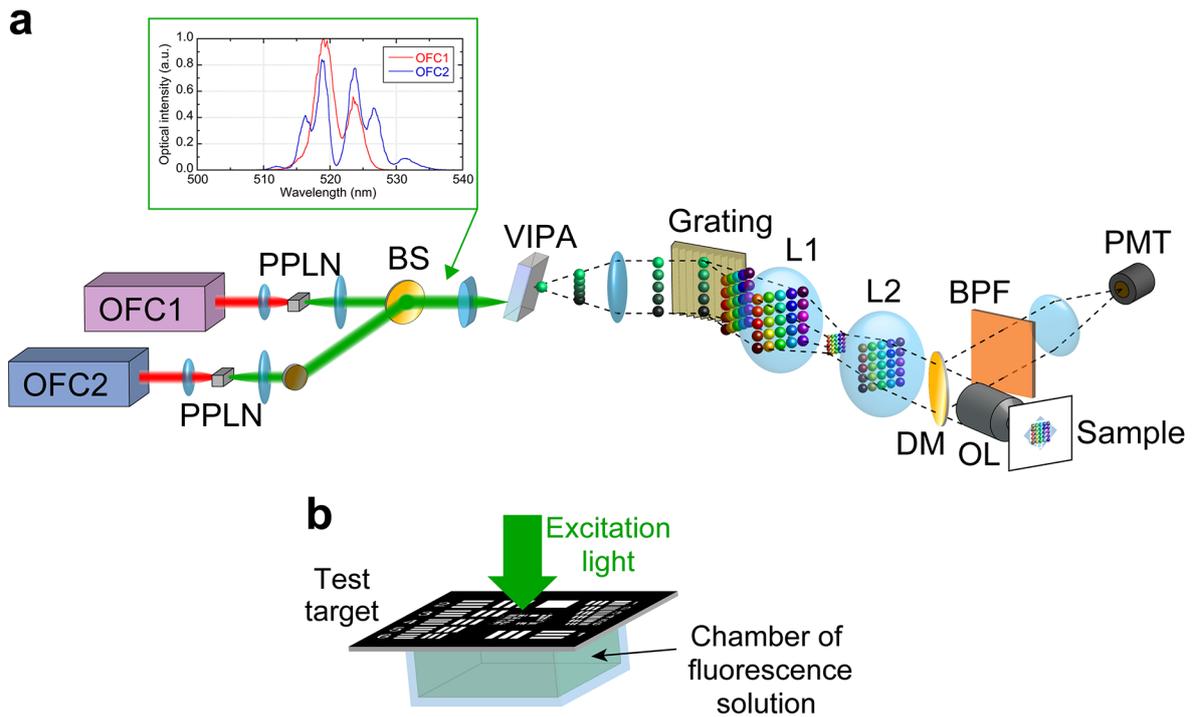

**Figure 1. a**, Experimental setup for f-DCM. OFC1 and OFC2, dual OFCs; PPLNs, chirped periodically-poled-lithium-niobate crystals; BS, beam splitter; VIPA, virtually imaged phased array; L1 and L2, lenses; DM, dichroic mirror; OL, objective lens; BPF, optical bandpass filter; PMT, photomultiplier. Inset shows optical spectra of OFC1 and OFC2 after converting into green light. **b**, Schematic drawing of the sample.

-30-

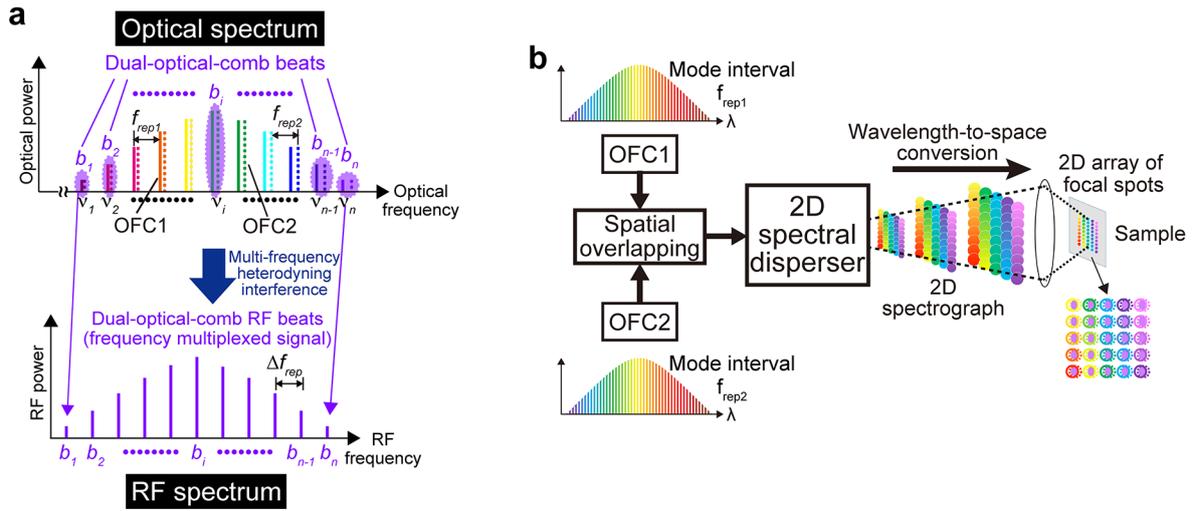

**Figure 2. a**, Dual-optical-comb beating between dual OFCs. **b**, 2D spectral mapping of dual OFCs.

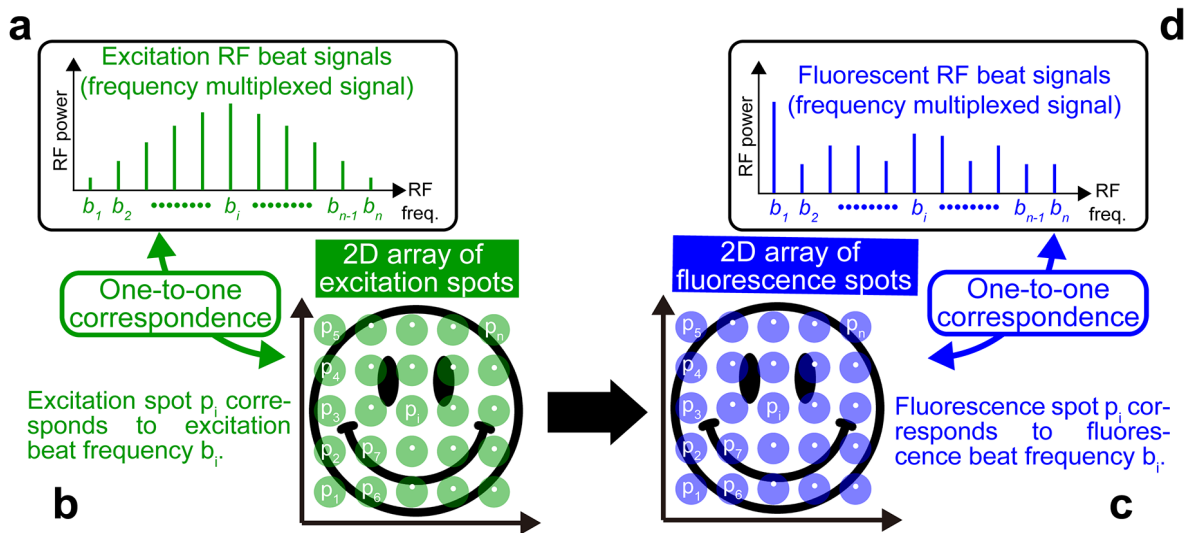

**Figure 3.** RF frequency multiplexing of excitation light: **a**, Frequency domain. **b**, Image domain. RF frequency multiplexing of fluorescence: **c**, Image domain. **d**, Frequency domain.

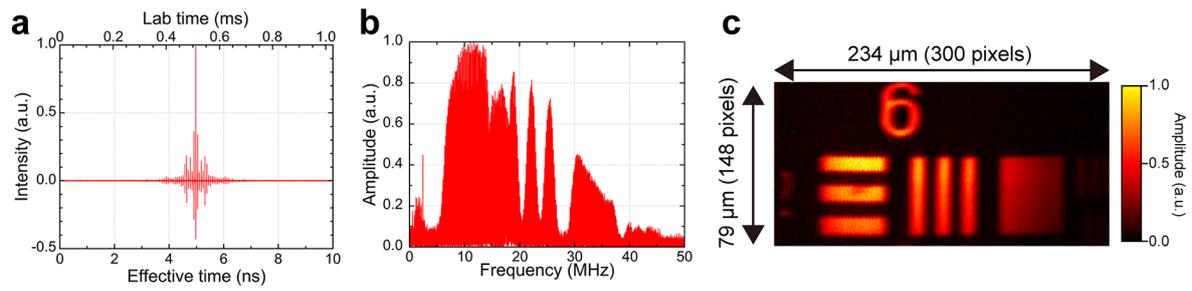

**Figure 4. a**, Temporal waveform. **b**, Amplitude spectrum of fluorescent RF comb modes. **c**, Fluorescence intensity image reconstructed from the amplitude spectrum of fluorescent RF comb modes.

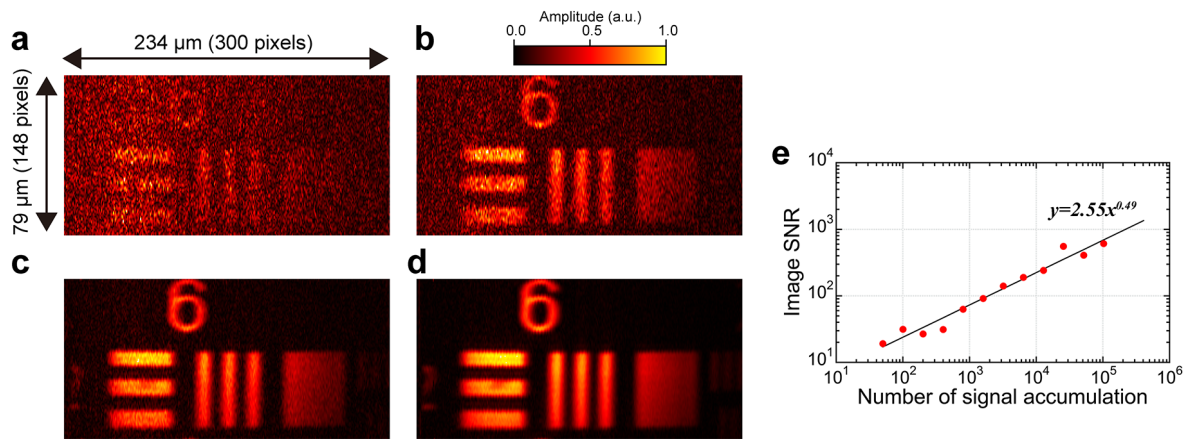

**Figure 5**. Change of fluorescence intensity image with respect to the number of accumulated signals: **a**, 100. **b**, 1,000. **c**, 10,000. **d**, 100,000. **e**, Relation between number of accumulated signals and image SNR.

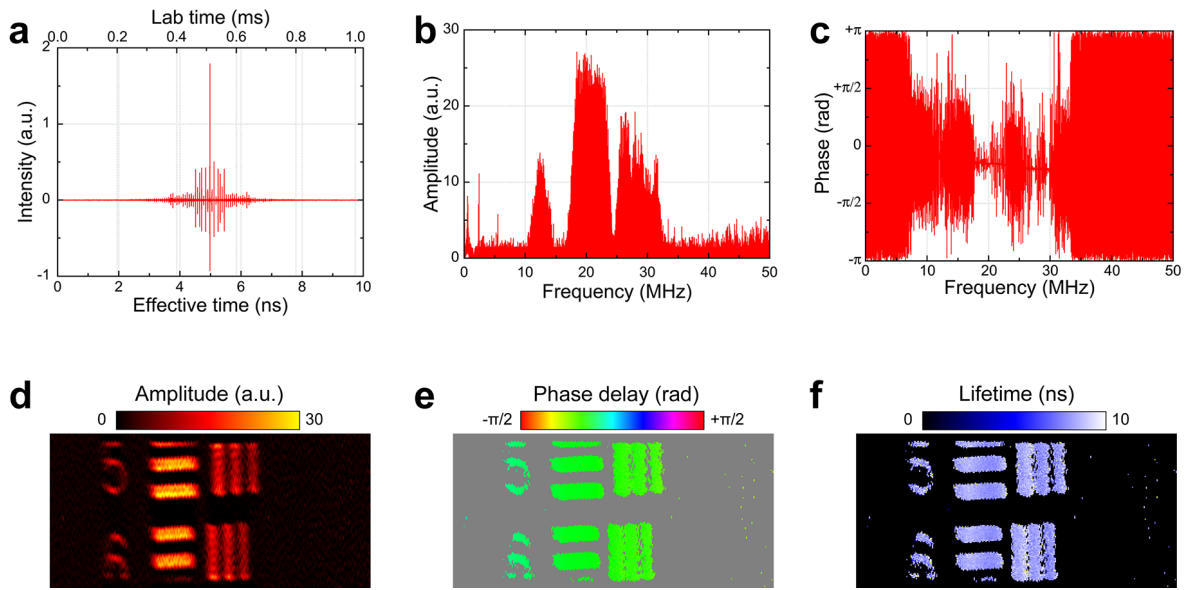

**Figure 6**. Fluorescent RF comb modes: **a**, Temporal waveform. **b**, Amplitude spectrum. **c**, Phase spectrum. **d**, Fluorescence intensity image. **e**, Fluorescence phase delay image. **f**, Fluorescence lifetime image.

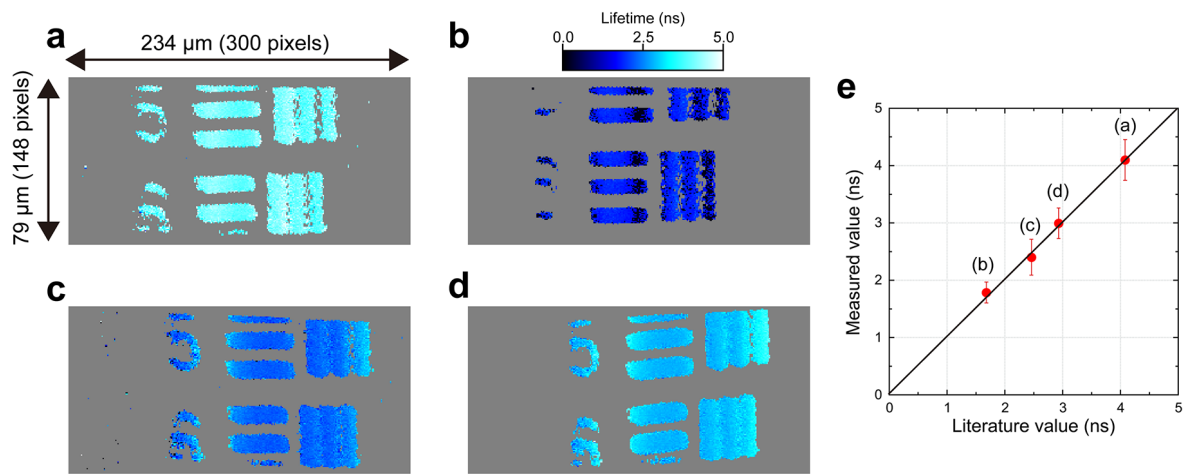

**Figure 7.** Fluorescence lifetime images with different fluorescent lifetimes: **a**, Rhodamine 6G aqueous solution. **b**, Rhodamine B aqueous solution. **c**, Rhodamine B methanol solution. **d**, Rhodamine B ethanol solution. **e**, Comparison of fluorescence lifetime between literature values and measured values.

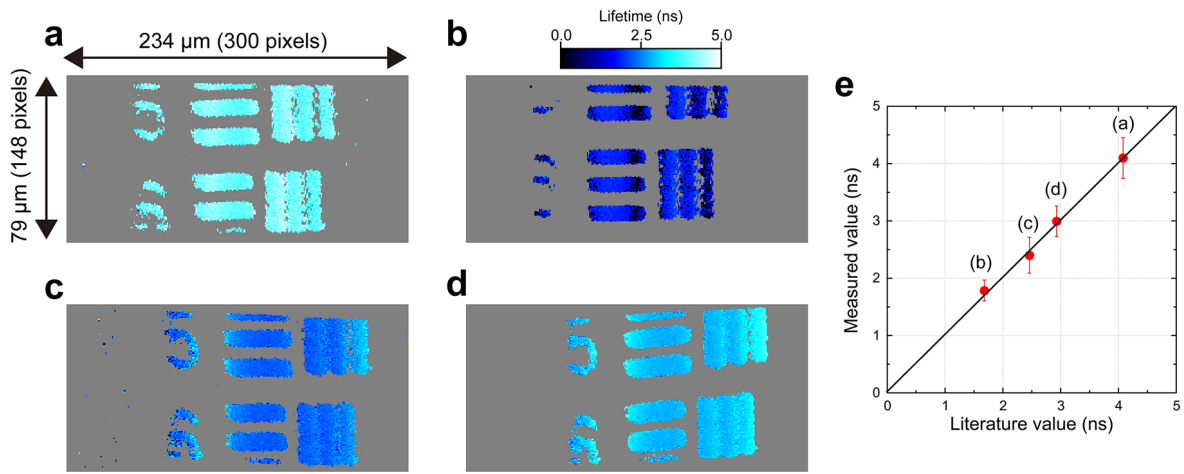

**Figure 8**. Phase measurement method for measuring fluorescence lifetime. **a**, Sinusoidal excitation light and the corresponding fluorescence. **b**, Phase delay between excitation light and fluorescence.

Table 1 Characteristics of prepared fluorescent solutions[34,35]

|  | Molar concentration (μM) | Fluorescence lifetime (ns) |
|---|---|---|
| Rhodamine 6G aqueous solution | 300 | 4.08 |
| Rhodamine B aqueous solution | 300 | 1.68 |
| Rhodamine B methanol solution | 300 | 2.46 |
| Rhodamine B ethanol solution | 300 | 2.93 |